\documentclass[twocolumn]{aastex63}

\newcommand{\chn}{{\it Chandra}}

\usepackage{graphicx}
\usepackage{longtable}
\usepackage{mdwlist}

\usepackage{natbib}
\usepackage{multirow}
\usepackage{upgreek}
\usepackage{wasysym}
\usepackage{txfonts}
\usepackage{gensymb}
\usepackage{morefloats}
\usepackage{threeparttable}
\usepackage{rotating}
\usepackage{graphicx}

\usepackage{lineno}
\shorttitle{The \chn\ 3CR extragalactic survey at high redshift}
\shortauthors{Jimenez Gallardo et al.}

\begin{document}

\title{Completing the 3CR Chandra snapshot survey: \\ extragalactic radio sources at high redshift} 

\correspondingauthor{Ana Jimenez-Gallardo}
\email{ana.jimenezgallardo@edu.unito.it}

\author[0000-0003-4413-7722]{A. Jimenez-Gallardo}
\affiliation{Dipartimento di Fisica, Universit\`a degli Studi di Torino, via Pietro Giuria 1, I-10125 Torino, Italy}
\affiliation{Istituto Nazionale di Fisica Nucleare, Sezione di Torino, I-10125 Torino, Italy}
\affiliation{INAF-Osservatorio Astrofisico di Torino, via Osservatorio 20, 10025 Pino Torinese, Italy}

\author[0000-0002-1704-9850]{F. Massaro}
\affiliation{Dipartimento di Fisica, Universit\`a degli Studi di Torino, via Pietro Giuria 1, I-10125 Torino, Italy}
\affiliation{Istituto Nazionale di Fisica Nucleare, Sezione di Torino, I-10125 Torino, Italy}
\affiliation{INAF-Osservatorio Astrofisico di Torino, via Osservatorio 20, 10025 Pino Torinese, Italy}
\affiliation{Consorzio Interuniversitario per la Fisica Spaziale, via Pietro Giuria 1, I-10125 Torino, Italy}

\author{M. A. Prieto}
\affiliation{Departamento de Astrof\'isica, Universidad de La Laguna, E-38206 La Laguna, Tenerife, Spain}
\affiliation{Instituto de Astrof\'isica de Canarias (IAC), E-38200 La Laguna, Tenerife, Spain}

\author[0000-0001-8382-3229]{V. Missaglia}
\affiliation{Dipartimento di Fisica, Universit\`a degli Studi di Torino, via Pietro Giuria 1, I-10125 Torino, Italy}
\affiliation{Istituto Nazionale di Fisica Nucleare, Sezione di Torino, I-10125 Torino, Italy}
\affiliation{INAF-Osservatorio Astrofisico di Torino, via Osservatorio 20, 10025 Pino Torinese, Italy}

\author[0000-0003-1619-3479]{C. Stuardi}
\affiliation{ADipartimento di Fisica e Astronomia, Universit\`a di Bologna, via Piero Gobetti 93/2, I-40129 Bologna, Italy}
\affiliation{Istituto di Radioastronomia, INAF, via Gobetti 101, 40129, Bologna, Italy}

\author[0000-0002-5646-2410]{A. Paggi}
\affiliation{Dipartimento di Fisica, Universit\`a degli Studi di Torino, via Pietro Giuria 1, I-10125 Torino, Italy}
\affiliation{Istituto Nazionale di Fisica Nucleare, Sezione di Torino, I-10125 Torino, Italy}
\affiliation{INAF-Osservatorio Astrofisico di Torino, via Osservatorio 20, 10025 Pino Torinese, Italy}

\author[0000-0001-5742-5980]{F. Ricci}
\affiliation{Instituto de Astrof\'isica and Centro de Astroingenier\'ia, Facultad de F\'isica, Pontificia Universidad Cat\'olica de Chile, Casilla 306, Santiago 22, Chile}

\author[0000-0002-0765-0511]{R. P. Kraft}
\affiliation{Center for Astrophysics $|$ Harvard \& Smithsonian, 60 Garden Street, Cambridge, MA 02138, USA}

\author[0000-0003-0995-5201]{E. Liuzzo}
\affiliation{Istituto di Radioastronomia, INAF, via Gobetti 101, 40129, Bologna, Italy}

\author[0000-0002-5445-5401]{G. R. Tremblay}
\affiliation{Center for Astrophysics $|$ Harvard \& Smithsonian, 60 Garden Street, Cambridge, MA 02138, USA}

\author{S. A. Baum}
\affiliation{University of Manitoba,  Dept. of Physics and Astronomy, Winnipeg, MB R3T 2N2, Canada}
\affiliation{Center for Imaging Science, Rochester Institute of Technology, 84 Lomb Memorial Dr., Rochester, NY 14623, USA}

\author[0000-0001-6421-054X]{C. P. O'Dea}
\affiliation{University of Manitoba,  Dept. of Physics and Astronomy, Winnipeg, MB R3T 2N2, Canada}
\affiliation{School of Physics \& Astronomy, Rochester Institute of Technology, 84 Lomb Memorial Dr., Rochester, NY 14623, USA}

\author[0000-0003-1809-2364]{B. J. Wilkes}
\affiliation{Center for Astrophysics $|$ Harvard \& Smithsonian, 60 Garden Street, Cambridge, MA 02138, USA}

\author{J. Kuraszkiewicz}
\affiliation{Center for Astrophysics $|$ Harvard \& Smithsonian, 60 Garden Street, Cambridge, MA 02138, USA}

\author[0000-0002-9478-1682]{W. R. Forman}
\affiliation{Center for Astrophysics $|$ Harvard \& Smithsonian, 60 Garden Street, Cambridge, MA 02138, USA}

\author{D. E. Harris}
\altaffiliation{Dan Harris passed away on December 6th, 2015. His career spanned much of the history of radio and X-ray astronomy. His passion, insight, and contributions will always be remembered. A significant fraction of this work is one of his last efforts.}
\affiliation{Center for Astrophysics $|$ Harvard \& Smithsonian, 60 Garden Street, Cambridge, MA 02138, USA}

\begin{abstract} 

We present the analysis of nine radio sources belonging to the Third Cambridge Revised catalog (3CR) observed with {\it Chandra} during Cycle 20 in the redshift range between 1.5 and 2.5. This study completes the 3CR $Chandra$ Snapshot Survey thus guaranteeing the X-ray coverage of all 3CR sources identified to date. This sample lists two compact steep spectrum sources, four radio galaxies and three quasars. We detected X-ray emission from all nuclei, with the only exception of 3C\,326.1 and 3C\,454.1 and from radio lobes in 6 out of 9 sources at level of confidence larger than $\sim$5$\sigma$. We measured X-ray fluxes and luminosities for all nuclei and lobes in the soft (0.5 - 1 keV), medium (1 - 2 keV) and hard (2 - 7 keV) X-ray bands. Since the discovered X-ray extended emission is spatially coincident with the radio structure in all cases, its origin could be due to Inverse Compton scattering of the Cosmic Microwave Background (IC/CMB) occurring in radio lobes. 
\end{abstract}

\keywords{galaxies: active --- X-rays: general --- radio continuum: galaxies}

\section{Introduction}
\label{sec:intro}
One of the most valuable catalogs of radio-loud active galactic nuclei (AGNs) is the Third Cambridge catalog of radio sources, originally built at 159 MHz (3C; \citealt{Edge1959} and \citealt{Bennet62}) and revised afterward at 178 MHz (see e.g., 3CR and 3CRR by \citealt{Spinrad1985} and \citealt{Laing1983}, respectively). The extragalactic fraction of the 3C catalog is a premiere sample of powerful radio galaxies and quasars spanning a wide range of radio powers and redshifts, from very nearby sources at $z\sim$0.001 up to more distant quasars lying at $z\sim$2.5, all radio selected at 178 MHz in the 3CR release (\citealt{Spinrad1985}). It lists both radio galaxies (RGs; i.e., narrow emission line objects as defined by \citealt{Jackson1990}) and quasars (QSOs; i.e., broad emission line objects as defined by \citealt{Jackson1990}), as well as edge-darkened (i.e., FR\,I) and edge-brightened (i.e., FR\,II) radio sources, according to the FR classification scheme (\citealt{Fanaroff1974}), and, both, high excitation (HERG) and low excitation (LERG) radio sources, classified on the basis of their optical spectra. This last classification scheme was first introduced by \citet{Hine1979} and was used since then by several authors to classify radio sources (e.g., \citealt{Laing1994}, \citealt{Buttiglione2010}, \citealt{Best2012} and \citealt{Baldi2019}).


This catalog has been extensively studied with the goal of exploring feedback processes occurring between radio galaxies and their environments (see, e.g., \citealt{Fabian2012} and \citealt{Kraft2012}) as well as investigating their nuclear emission and the emission from X-ray counterparts of their extended radio structures (i.e., jet knots, hotspots, lobes; see e.g., \citealt{Croston2005}, \citealt{Ineson2013,Ineson2015,Ineson2017} and \citealt{Mingo2017}). Thus, a vast suite of observations were performed in the last decades for the 3CR catalog in the radio (see e.g., \citealt{Law1995} and \citealt{Giovannini2005}), infrared (see e.g., \citealt{Baldi2010}, \citealt{Werner2012}, \citealt{Dicken2014} and \citealt{Ghaffari2017}), optical (see e.g., \citealt{Hiltner1991}, \citealt{deKoff1996}, \citealt{Chiaberge2000}, \citealt{Buttiglione2009,Buttiglione2011} and \citealt{Tremblay2009}), X-ray (see e.g. \citealt{Prieto1996}) and gamma (\citealt{Torresi2018}) bands, using all major observing facilities, such as the Very Large Array (VLA), Spitzer and Herschel, Hubble Space Telescope (HST), the ROentgen SATellite (ROSAT) and Fermi satellites, to name a few examples. 

Then to complete the X-ray coverage of the 3CR catalog, in 2007, we embarked on the 3CR $Chandra$ Snapshot Survey, with the goal of obtaining snapshot observations (i.e., exposure time of the order of 20 ksec) for the entire catalog. Before this project, only 150 extragalactic radio sources out of 298 listed in the 3CR catalog were already present in the $Chandra$ archive (see e.g., \citealt{Evans2006} and \citealt{Hardcastle2006a}). Thus in $Chandra$ Cycle 9, we started observing all 3CR sources at $z<0.3$ lacking X-ray observations. 29 targets were observed during $Chandra$ Cycle 9 (\citealt{Massaro2010}) and the remaining 27 during Cycle 12 (\citealt{Massaro2012}), with nominal exposure times of 8 ksec each. We then continued requesting radio sources at $z>0.3$ (increasing the exposure time, taking into account the decrease in the low-energy effective area due to the build-up of contaminant on the ACIS-S filter) and observed 19 radio galaxies in $Chandra$ Cycle 13 up to $z=0.5$ (\citealt{Massaro2013}); 23 more targets in Cycle 15 (\citealt{Massaro2018}) and completed all 3CR observations up to $z=1.5$ in Cycle 17 (16 more sources, \citealt{Stuardi2018}). 


Here we present the last paper in our survey reporting $Chandra$ observations of the remaining 9 3CR sources, completing observations of all identified 3CR sources up to $z=$ 2.5, namely 3C\,239, 3C\,249,\,3C 257, 3C\,280.1, 3C\,322, 3C\,326.1, 3C\,418, 3C\,454 and 3C\,454.1. These are the highest redshift (from 1.554 to 2.474) sources listed in the 3CR $Chandra$ Snapshot Survey. Observations in the 3CR $Chandra$ Snapshot Survey have inspired several X-ray follow up observations of different targets of the survey (see e.g., \citealt{Massaro2009}, \citealt{Hardcastle2010,Hardcastle2012}, \citealt{Balmaverde2012}, \citealt{Orienti2012}, \citealt{Ricci2018} and \citealt{Paggi2020}).  Furthermore, this survey led to several discoveries of X-ray counterparts of radio jet knots and hotspots (see e.g., \citealt{Massaro2015}) and of extended X-ray emission around these radio sources (\citealt{Massaro2018}, \citealt{Stuardi2018} and \citealt{Jimenez2020}).
 
However, during our investigation, we also discovered that the 3CR catalog lists 25 targets that are still unidentified (\citealt{Massaro2015} and \citealt{Maselli2016}), lacking an association with an optical counterpart and thus also being unclassified. There is a high chance that these 3CR unidentified sources are i) high redshift quasars, ii) highly obscured radio-loud active galaxies, thus very different from the rest of the 3CR catalog, deserving a deeper analysis, or iii) LERGs, and thus lacking radiatively efficient AGN signatures in the optical and X-rays (see \citealt{Hardcastle2009} and \citealt{Maselli2016}). A $Chandra$ campaign is currently ongoing to observe the first 9 unidentified targets and results will be presented in a forthcoming paper (Massaro et al. 2020 in prep.).

As in previous data papers, the current manuscript is organized as follows. A brief overview of the data reduction procedures, uniformly applied to all previous analyses, is given in \S~\ref{sec:obs} while results are described in \S~\ref{sec:results}. X-ray images  of the 9 3CR radio galaxies are shown \S~\ref{sec:results}, with radio contours overlaid. Then, \S~\ref{sec:summary} is devoted to our summary, conclusions and a brief overview of future perspectives. 

Unless otherwise stated, we adopted cgs units for numerical results and assumed the same flat cosmology with $H_0=69.6$ km s$^{-1}$ Mpc$^{-1}$, $\Omega_{M}=0.286$ and $\Omega_{\Lambda}=0.714$ \citep{bennett14}, previously adopted in earlier papers in this series. Spectral indices, $\alpha$, are defined by flux density, S$_{\nu}\propto\nu^{-\alpha}$. Average background level of the X-ray images is $\sim0.02$ photons pixel$^{-1}$.

\section{Data reduction and data analysis}
\label{sec:obs}

Here we present an overview of the data reduction and analysis procedures performed on \chn\ observations. To create a uniform \chn\ database of 3CR sources, procedures performed are the same as those in our previous papers and, therefore, details about them can be found in \citet{Massaro2010,Massaro2011,Massaro2012,Massaro2013,Massaro2015,Massaro2018} and \citet{Stuardi2018}.

Data reduction was done following the standard procedure described in the $Chandra$ Interactive Analysis of Observations (CIAO; \citealt{Fruscione2006}) threads\footnote{http://cxc.harvard.edu/ciao/threads/}. In particular, we used CIAO v4.10 with the \chn\ Calibration Database (CALDB) version 4.8.4.1. Level 2 event files were created using the CIAO task \textsc{chandra\_repro}. Astrometric registration was carried out by aligning the radio core position, when detected, with its X-ray position (following the same procedure described in \citealt{Massaro2011}). In our case, only three sources, namely, 3C\,280.1, 3C\,418 and 3C\,454 were registered since radio nuclei could not be detected using radio maps available for the remaining sources. For sources registered, the total shift was of the order of 0.5 arcsec or less, which corresponds to $\sim4$ kpc at redshifts 1.5 - 2.5, consistent with previous campaign papers.

Flux maps were created by taking into account the exposure time and the effective area. We obtained maps in the X-ray energy ranges: 0.5 -- 1 keV (soft), 1 -- 2 keV (medium) and 2 -- 7 keV (hard) by using monochromatic exposure maps set to the nominal energies of 0.75, 1.4, and 4 keV for the soft, medium and hard bands, respectively. All flux maps were converted from units of counts s$^{-1}$ cm$^{-2}$ to cgs units by multiplying each event by the nominal energy of each band, assuming that every event in the same band has the same energy (see \citealt{Massaro2015}). 

We defined the position of X-ray nuclei and lobes using radio maps. However, when the core was not detected in radio, the position of the X-ray nucleus was defined as the emission peak in the 5 - 7 keV band (in contrast to the 0.5 - 7 keV band used for images and for the rest of the analysis) since we expect nuclei to have hard spectra in the X-ray band. The positions found are consistent (within 2\arcsec) with those of the host galaxy in the Panoramic Survey Telescope and Rapid Response System (Pan-STARRS\footnote{https://catalogs.mast.stsci.edu/panstarrs/}; \citealt{Chambers2016}) and those reported in \citet{Spinrad1985}. Then, observed X-ray fluxes were measured using circular regions of 2\arcsec\ radii centered on the X-ray nuclei and, for lobes, circular regions containing the radio emission at more than three times the root mean square (rms) level of the background in the radio maps and centered on the peak of the radio emission. Background regions were chosen as circular regions at least as big as the radio sources and located on the same charge-coupled device (CCD) chip, far enough from the radio galaxy (i.e., at least a few tens of arcsec) to avoid the smearing of the point-spread function (PSF) on CCD borders and contamination from the source. Average background level is $\sim$0.02 photons pixel$^{-1}$.

Fluxes were measured in each energy band and region using funtools\footnote{http://www.cfa.harvard.edu/$\sim$john/funtools} and assuming a flat energy response in each band as in our previous analyses (see \citealt{Massaro2015}). Uncertainties were computed assuming Poisson statistics (i.e., square root of the number of photons) in the source and background regions. X-ray fluxes were not corrected for the Galactic absorption, as in our previous analyses (see e.g., \citealt{Massaro2015}), since we would need to assume a given photon index. X-ray luminosities were not corrected for neutral hydrogen column densities (either galactic or intrinsic), as in our previous analyses. Results for nuclei are reported in Table~\ref{tab:cores} while lobes fluxes are given in Table~\ref{tab:features}. We indicate here as lobes X-ray counterparts all diffuse emission detected in the 0.5-7 keV energy range consistent/associated with extended radio structures, since at these high redshifts radio maps available do not have sufficient resolution to distinguish/identify hotspots. The lobes are labeled as a combination of one letter indicating their orientation and one number indicating their distance from the X-ray nuclei in arcseconds.

The \chn\ native pixel size for the ACIS instrument is 0\arcsec.492, while the half-energy radius is 0\arcsec.418, so the data are under-sampled\footnote{https://cxc.harvard.edu/proposer/POG/pdf/MPOG.pdf}. To recover the resolution, images were regridded to 1/2, 1/4, or 1/8 of the native ACIS pixel size, according to the angular sizes of radio sources and to their number of counts (see also \citealt{Massaro2012,Massaro2013} for more details).

As in previous analyses, a comparison between radio and X-ray images was carried out. This allowed us to verify the presence of X-ray counterparts of radio structures and to identify those sources that present diffuse X-ray emission in the soft band that is extended beyond the position of the radio lobes. This diffuse emission could be either due to Inverse Compton scattering of seed photons arising from the Cosmic Microwave Background (IC/CMB) or X-ray emission arising from intracluster medium (ICM) in galaxy clusters or a combination of both processes (see e.g., \citealt{Croston2005}). We expect this diffuse X-ray emission to peak in the soft band, regardless of its origin. To obtain the detection significance of the diffuse X-ray emission, we assumed Poisson statistics as for nuclei and lobes, using also the same background regions. Emission regions were chosen as circles centered in the X-ray nuclei and including all the radio emission above a 3 rms level and excluding the regions chosen for the nuclei and the lobes.

Radio images were obtained from the VLA archive website\footnote{https://archive.nrao.edu/cgi-bin/nvas-pos.pl} as well as one of our colleague's personal website\footnote{http://www.slac.stanford.edu/~teddy/vla3cr/vla3cr.html} and the Tata Institute of Fundamental Research (TIFR) Giant Metrewave Radio Telescope (GMRT) Sky Survey (TGSS\footnote{http://tgssadr.strw.leidenuniv.nl/doku.php}). Image parameters for each radio observation used are given in the figure captions of \S~\ref{sec:results}.

An additional difference with respect to previous survey papers is the lack of spectral analysis. Since all sources in this sample are at high redshift, using only snapshot observations meant that the total number of photons received from each source was too small to carry out this kind of analysis.

\section{Results}
\label{sec:results}

\subsection{General}

We detected X-ray emission in the full band (0.5 - 7 keV) above a 3$\sigma$ significance level for all the nuclei in our sample with the only exception of 3C\,326.1 and 3C\,454.1. Background-subtracted number of photons within a circle of $r=2$ \arcsec\, as well as the nuclear fluxes in the soft (0.5 - 1 keV), medium (1 - 2 keV), hard (2 - 7 keV) and full (0.5 - 7 keV) bands and the X-ray luminosity for each source are shown in Table \ref{tab:cores}. In contrast with previous analyses (see, e.g., \citealt{Massaro2012}), we do not report the extent ratio of the sources (i.e., the ratio of photons in the 2\arcsec\ circle and those in an annulus between 2\arcsec\ and 10\arcsec) since nearly all sources in our sample are enclosed within 10\arcsec\ from their X-ray nucleus.



We detected 9 lobes above a 3$\sigma$ confidence level in 6 of the sources in our sample. Sizes, background-subtracted number of photons, X-ray fluxes and X-ray luminosities of the X-ray counterparts of radio lobes are listed in Table \ref{tab:features}.

X-ray detection of nuclei and lobes was carried out in the 0.5 - 7 keV band like in previous analyses for the 3CR $Chandra$ Snapshot Survey; while we used the 0.5 - 3 keV band to detect the diffuse X-ray emission, since we expect this emission to peak in the soft band, regardless of its origin. This diffuse emission was estimated excluding the regions selected for nuclei and lobes.

Three out of the 7 nuclei detected in X-rays in our sample are affected by pileup (see \citealt{Davis2001} for details about pileup), namely 3C\,280.1 (5\%), 3C\,418 (10\%) and 3C\,454 (3\%). Therefore, X-ray fluxes obtained for these sources are underestimated.

Out of the 9 sources in our sample, we only detected diffuse X-ray emission in the soft band along the radio axis and beyond the lobe radio emission observed at frequencies in the order of GHz in 3C\,249, with a 2$\sigma$ confidence level. This diffuse emission is shown in the left panel of Fig. \ref{fig:3c249}, where the X-ray emission in the 0.5 - 3 keV band with 150 MHz TGSS radio contours overlaid is shown. Additionally, only two out of the 9 selected targets, namely 3C\,249 and 3C\,322 presented extended morphology in TGSS (see left panels of Figs. \ref{fig:3c249} and \ref{fig:3c322}), while the rest are unresolved.

As carried out for the analyses of 3CR sources at low redshifts (i.e., below 0.3) as well as in works such as the ones by \citet{Hardcastle2006b,Hardcastle2009}, \citet{Mingo2014,Mingo2016}, \citet{Panessa2016}, \citet{Ursini2018} and \citet{Butler2019} on similar samples; we performed spectral simulations with {\sc sherpa}, using a model comprising Galactic absorption (fixed to the values reported in Table \ref{tab:log} for each source), a power-law with slope fixed to 1.8, and intrinsic absorption with column density N$_{H,int}$ at the source redshift $z$. We computed the ranges of the hardness rations $HR$ as the ratio between the fluxes in the soft (0.5 - 3 keV) and the hard band (3 - 7 keV) and, from those, we derived the range of intrinsic absorbing column density N$_{H,int}$ as reported in Table \ref{tab:cores} for all our detected nuclei, as well as the ranges of X-ray luminosities in the 0.5 - 7 keV band without taking into account any absorption ($L_X$) and from the model with Galactic and intrinsic absorptions ($L_{X,obs}$).

Moreover, for three sources (namely, 3CR\,280.1, 3CR\,418 and 3CR\,454), the number of detected photons for their nuclei allowed us to carry out detailed spectral analyses as we did for relatively bright 3CR sources at lower redshifts observed during our snapshot survey. All spectral fits in these cases were consistent with simple power-law with galactic absorption, where the photon index is consistent with that of typical AGNs (i.e., 1.8) within 3$\sigma$ and, in agreement with the HR analysis, the column density from intrinsic absorption is negligible with respect to the Galactic one and/or unconstrained. A summary of these results can be seen in Table \ref{tab:cores}.

\subsection{Source details}

\textit{3C\,239}.
This is a FR II radio galaxy at $z=1.781$ (\citealt{Laing1983}) optically identified as a high-excitation radio galaxy (HERG; \citealt{Jackson1997}). According to \citet{Hammer1990}, this galaxy lies in a crowded field where there are two different populations of galaxies, one at the same redshift as 3C\,239 and one in the foreground. They suggested that the first population belongs to a galaxy cluster surrounding 3C\,239 which could be gravitationally lensed by the foreground population. The last statement was supported by \citet{Best1997} since they claim that a foreground galaxy within 5\arcsec\ of the host of 3C\,239, next to the eastern lobe, could cause the presence of a double hotspot and arc-like structures in that lobe. The $HST$ images (\citealt{Best1997}) show a bright galaxy with two emission regions aligned with the radio axis and a string of emission to the south-east misaligned to the radio axis by $\sim 45\degree$, as well as two faint tidal tails to the north and west of the galaxy, extending $\sim 40$ kpc. These features could be the result of past galaxy collisions.

\begin{figure}[h!]
\includegraphics[width=9.cm]{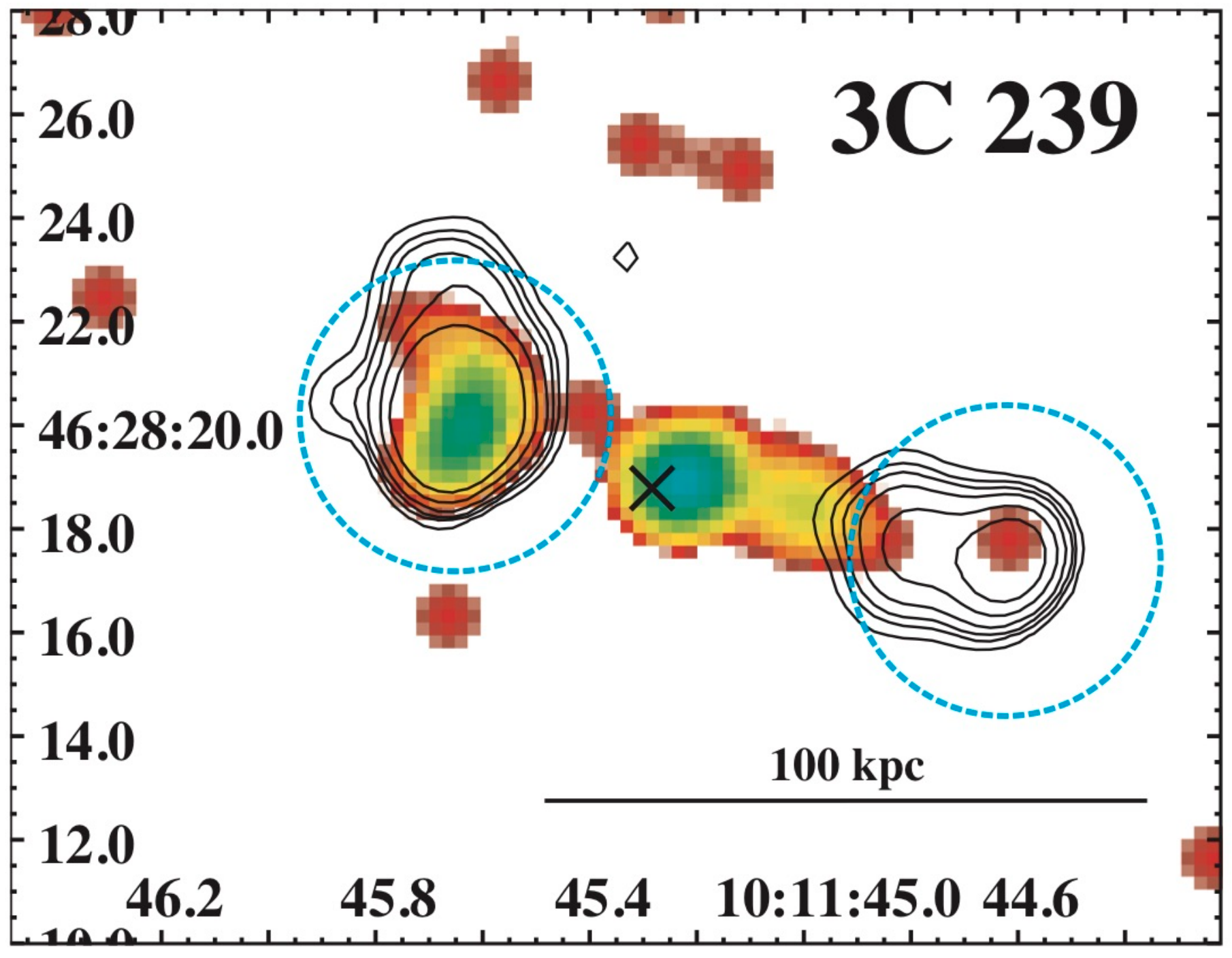}
\caption{0.5 - 7 keV $Chandra$ image with 1.8 GHz VLA contours overlaid. $Chandra$ image was rebinned by a factor of 1/2 and smoothed with a 6 pixel Gaussian kernel. Radio contours were drawn at levels of 10, 20, 30, 50, 100 and 200 times the rms. Radio map has a beam size of 1.3 \arcsec, which is equivalent to 11.2 kpc at the source redshift. The position of the X-ray nucleus (identified as the peak emission in the 5 - 7 keV band) is shown with a black X. Blue dashed circles correspond to regions used to define lobes.}
\label{fig:3c239L}
\end{figure}

Both the nucleus and the eastern lobe of 3C\,239 were detected in the X-rays (using the 0.5 - 7 keV band), using Poisson statistics (see Fig. \ref{fig:3c239L}). We could not find an archival radio map in which the radio core was detected, therefore, we considered the position of the X-ray nucleus that of the peak emission in the 5 - 7 keV band, in which the nucleus was detected at a 3$\sigma$ confidence level. However, we did not detect the presence of diffuse X-ray emission beyond the radio lobes that could indicate the presence of ICM from a galaxy cluster.

\begin{figure*}
    \centering
\includegraphics[width=18.cm]{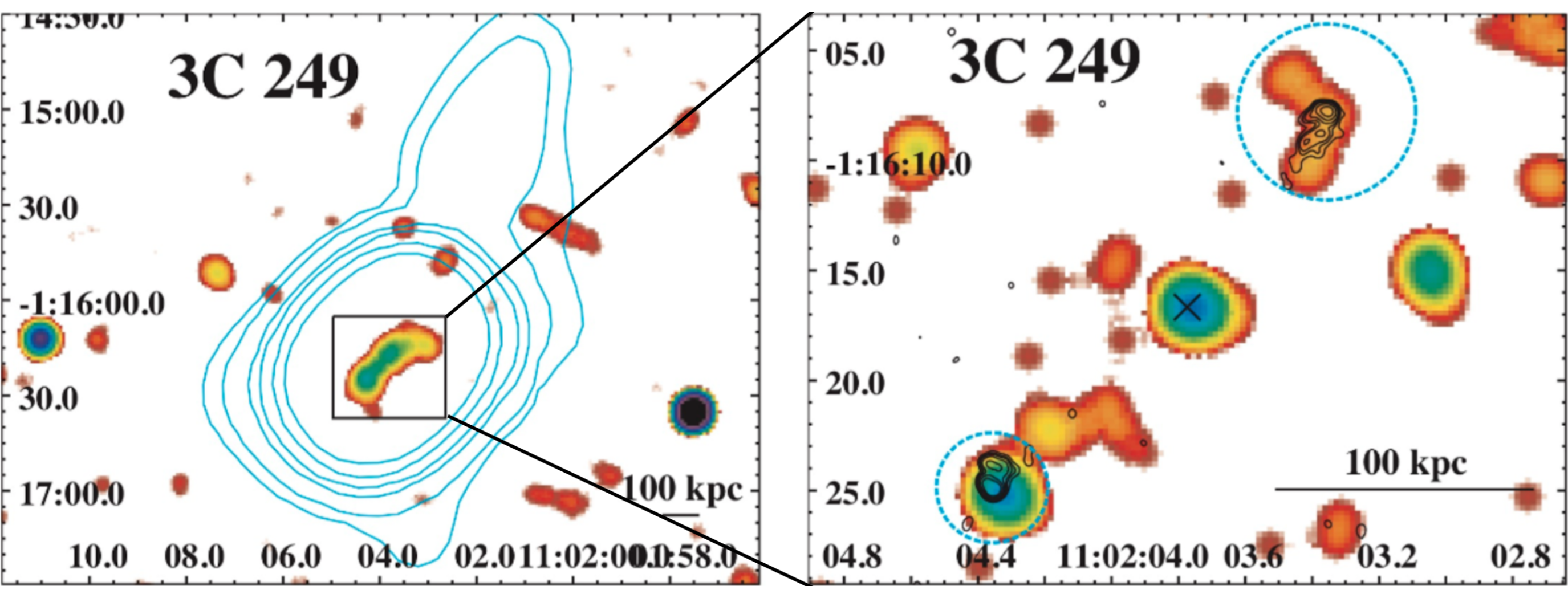}
\caption{Left: $Chandra$ 0.5 - 3 keV X-ray image with TGSS 150 MHz contours overlaid. Radio contours were drawn at 5, 10, 30, 50, 100 and 200 times the rms and the radio map has a beam size of 27 \arcsec, which is equivalent to 232.5 kpc at the source redshift. $Chandra$ image was binned by a factor 2 and smoothed with a 8 pixel Gaussian kernel. Right: 0.5 - 7 keV $Chandra$ image with 4.8 GHz VLA contours overlaid. $Chandra$ image was rebinned by a factor of 1/2 and smoothed with a 8 pixel Gaussian kernel. Radio contours were drawn at 10, 20, 30, 50, 100 and 200 times the rms and the radio map has a beam size of 0.5 \arcsec, which is equivalent to 4.3 kpc at the source redshift. The position of the X-ray nucleus is shown with a black X. Blue circles show regions used to identify lobes.}
\label{fig:3c249}
\end{figure*}

\textit{3C\,249}.
This source is a lobe-dominated quasar at $z=1.554$ (\citealt{Law1995}). It has not been studied in detail in the literature and, as for the previous source, we could not find a radio map in which the radio core was detected. However, we obtained the position of the nucleus using hard X-rays (see right panel of Fig. \ref{fig:3c249}), in which the nucleus was detected with a 2$\sigma$ confidence level. We were able to detect the X-ray nucleus and both lobes in the 0.5 - 7 keV band. Additionally, diffuse X-ray emission was found beyond the position of the radio lobes in the 0.5 - 3 keV band (with a 2$\sigma$ confidence level), as shown in the left panel of Fig. \ref{fig:3c249}.

\textit{3C\,257}. 
This is the highest redshift ($z=2.474$) radio galaxy in the 3C catalog. According to \citet{Hilbert2016}, the IR image shows a distorted morphology ($\sim2\arcsec$, which is equivalent to 16.5 kpc at the source redshift), which implies that the source could be undergoing a merging event. Additionally, it shows some optical emission which implies some ongoing star formation.

\begin{figure}[h!]
    \centering
\includegraphics[width=8.9cm]{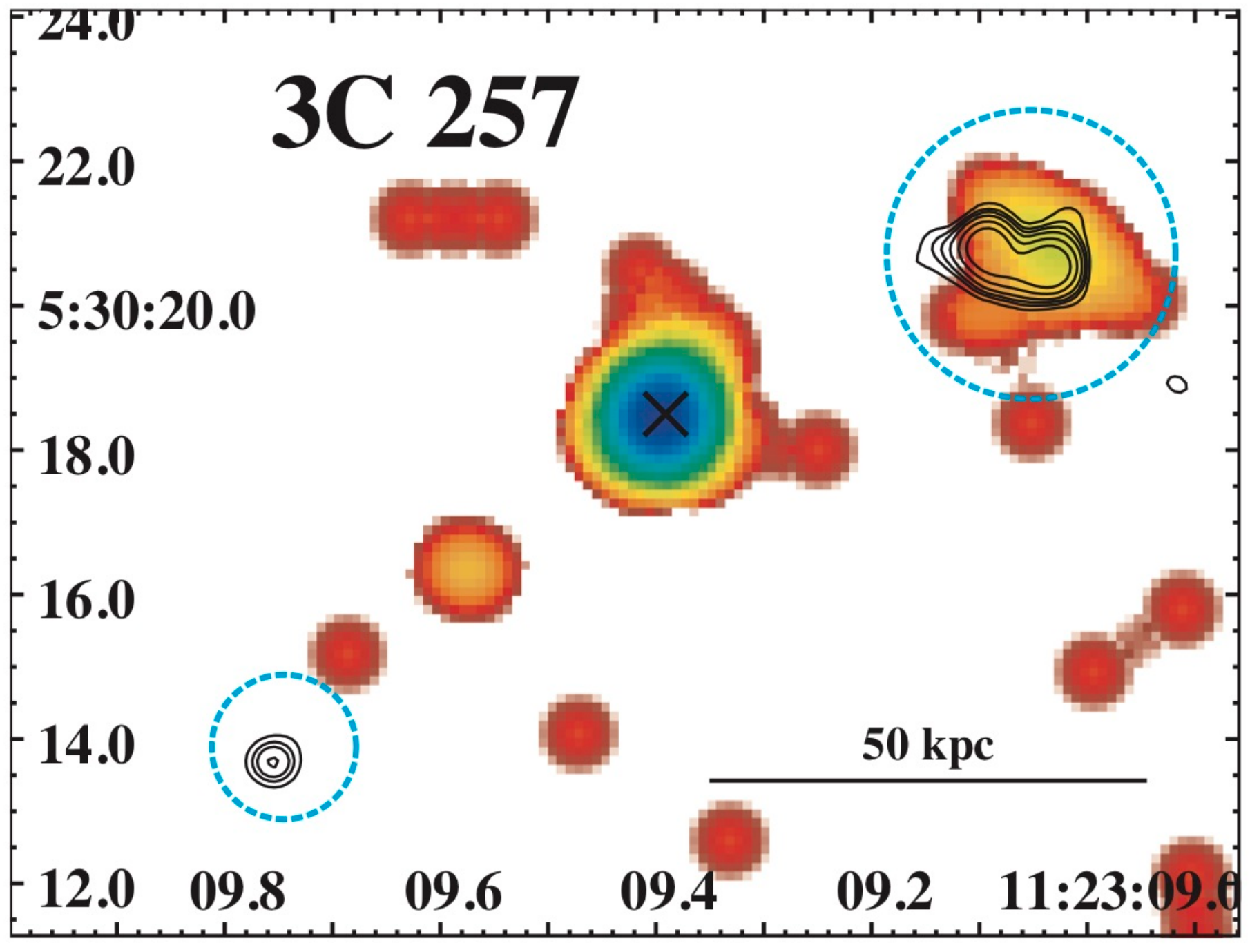}
\caption{0.5 - 7 keV $Chandra$ image with 4.8 GHz VLA contours overlaid. $Chandra$ image was rebinned by a factor of 1/4 and smoothed with a 8 pixel Gaussian kernel. Radio contours were drawn at levels of 10, 20, 30, 50, 100 and 200 times the rms. Radio map has a beam size of 0.5 \arcsec, which is equivalent to 4.1 kpc at the source redshift. The position of the X-ray nucleus is shown with a black X. Blue circles show regions used to identify lobes.}
\label{fig:3c257L}
\end{figure}

We defined the position of the X-ray nucleus using hard X-rays as in previous sources (in this band, the nucleus was detected with a 3$\sigma$ confidence level). In the $Chandra$ full band image, we detected the nucleus and the western lobe (see Fig. \ref{fig:3c257L}). We did not detect diffuse X-ray emission in the soft band only.

\textit{3C\,280.1}. 
This source is a lobe-dominated quasar at $z=1.667$ (\citealt{Laing1983} and \citealt{Veron2006}). The 5 GHz radio map in \citet{Lonsdale1993} shows a strong jet.

\begin{figure}[h!]
    \centering
\includegraphics[width=8.9cm]{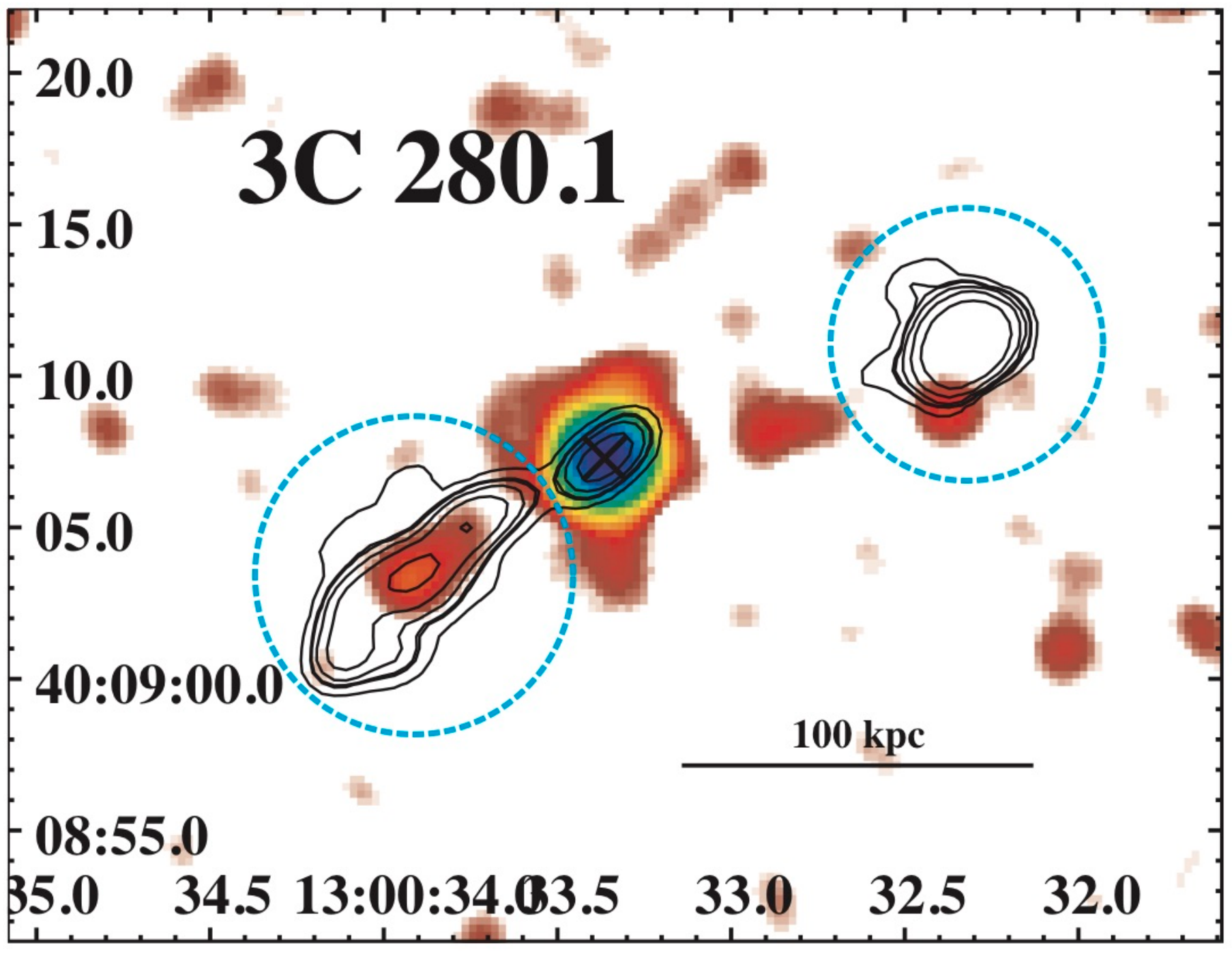}
\caption{0.5 - 7 keV $Chandra$ image with 1.8 GHz VLA contours overlaid. $Chandra$ image was rebinned by a factor of 1/2 and smoothed with a 6 pixel Gaussian kernel. Radio contours were drawn at levels of 10, 20, 30, 50, 100 and 200 times the rms. Radio map has a beam size of 1.5 \arcsec, which is equivalent to 12.9 kpc at the source redshift. The position of the radio core is shown with a black X. Blue circles show regions used to identify lobes.}
\label{fig:3c280.1L}
\end{figure}

In this case, we were able to define the core position using the radio emission (in this position, the X-ray nucleus was detected in the 5 - 7 keV band above a 5$\sigma$ confidence level). We detected the nucleus and both lobes in the 0.5 - 7 keV band (see Fig. \ref{fig:3c280.1L}). Furthermore, the eastern lobe of 3C\,280.1, which corresponds to the strong jet detected by \citet{Lonsdale1993} at radio frequencies, is the brightest lobe in our sample. The X-ray nucleus of 3C\,280.1 is affected by 5\% pileup. Lastly, we did not detect diffuse X-ray emission in the 0.5 - 3 keV band.

\textit{3C\,322}.
This is a FR II radio galaxy at $z=1.618$ (\citealt{Law1995}). In the IR, this source appears as an extended object ($\sim2\arcsec$, which is equivalent to 17 kpc at the source redshift) with a core elongated in the northwest-southeast direction (\citealt{Hilbert2016}). According to \citet{Kotyla2016}, it could be part of a galaxy cluster since there is an overdensity of galaxies on its field.

\begin{figure*}
    \centering
\includegraphics[width=18cm]{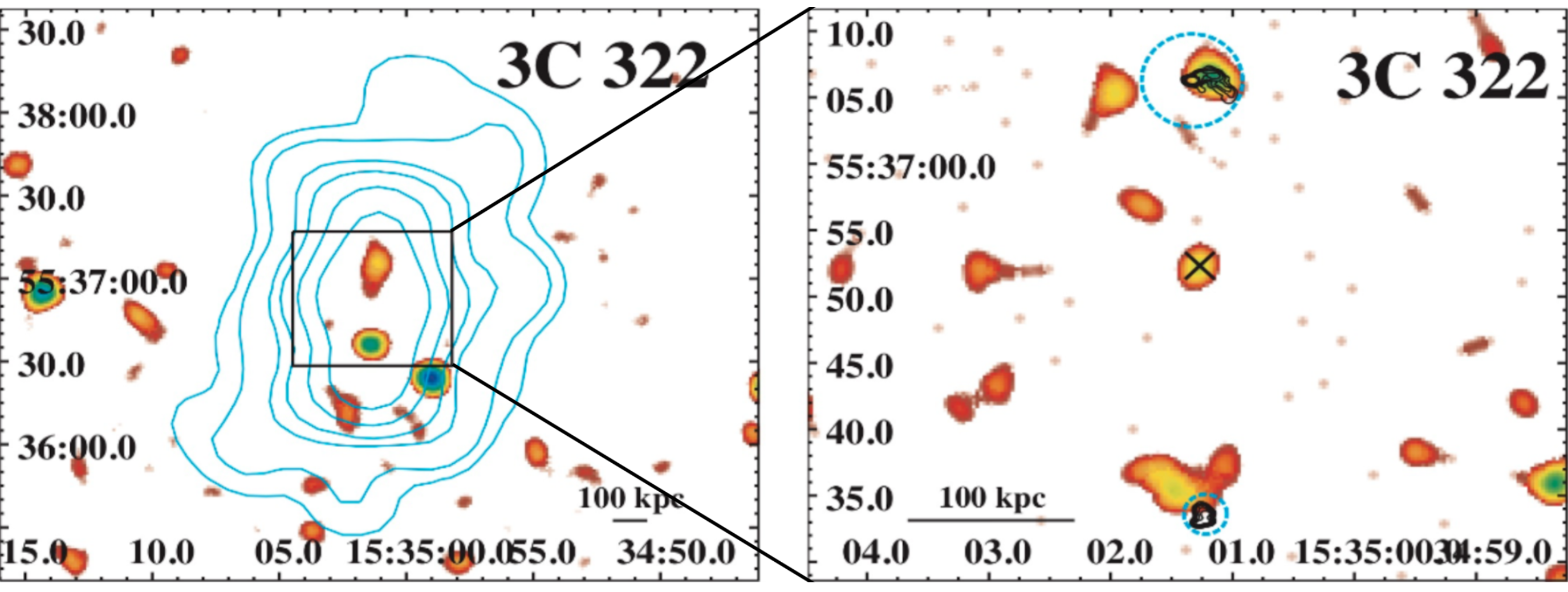}
\caption{Left: $Chandra$ 0.5 - 3 keV X-ray image with TGSS 150 MHz. Radio contours were drawn at 5, 10, 30, 50, 100 and 200 times the rms and the radio map has a beam size of 25 \arcsec, which is equivalent to 215.4 kpc at the source redshift. $Chandra$ image was binned by a factor 2 and smoothed with a 8 pixel Gaussian kernel. Right: 0.5 - 7 keV $Chandra$ image with 1.8 GHz VLA contours overlaid. $Chandra$ image was rebinned by a factor of 1/2 and smoothed with a 8 pixel Gaussian kernel. VLA radio contours were drawn at levels of 10, 20, 30, 50, 100 and 200 times the rms and the radio map has a beam size of 0.5 \arcsec, which is equivalent to 4.3 kpc at the source redshift. The position of the X-ray nucleus is shown with a black X. Blue circles show regions used to identify lobes.}
\label{fig:3c322}
\end{figure*}

In the full X-ray band, we detected the nucleus (whose position we found using the peak emission in hard X-rays with a 4$\sigma$ confidence level) and the north lobe (see right panel of Fig. \ref{fig:3c322}). We found no diffuse emission in the 0.5 - 3 keV band that could indicate the presence of ICM from a galaxy cluster (see left panel of Fig. \ref{fig:3c322}). 

\textit{3C\,326.1}. 
This source is a FR II radio galaxy at $z=1.825$. \citet {McCarthy1987} found that the host galaxy is surrounded by a giant $Ly\alpha$ cloud of $\sim 100$ kpc diameter, which could imply that it is a young and/or starforming galaxy. IR observations presented by \citet{Hilbert2016} show that the host is a dim galaxy, elongated in the east-west direction and surrounded by many small and irregular sources. \citet{Hilbert2016} also observed several clumps of UV emission due to, for example, star formation regions. Lastly, \citet{Kotyla2016} found an overdensity of sources in the field of 3C\,326.1, suggesting that it could be part of a galaxy cluster. 

\begin{figure}[h!]
    \centering
\includegraphics[width=8.9cm]{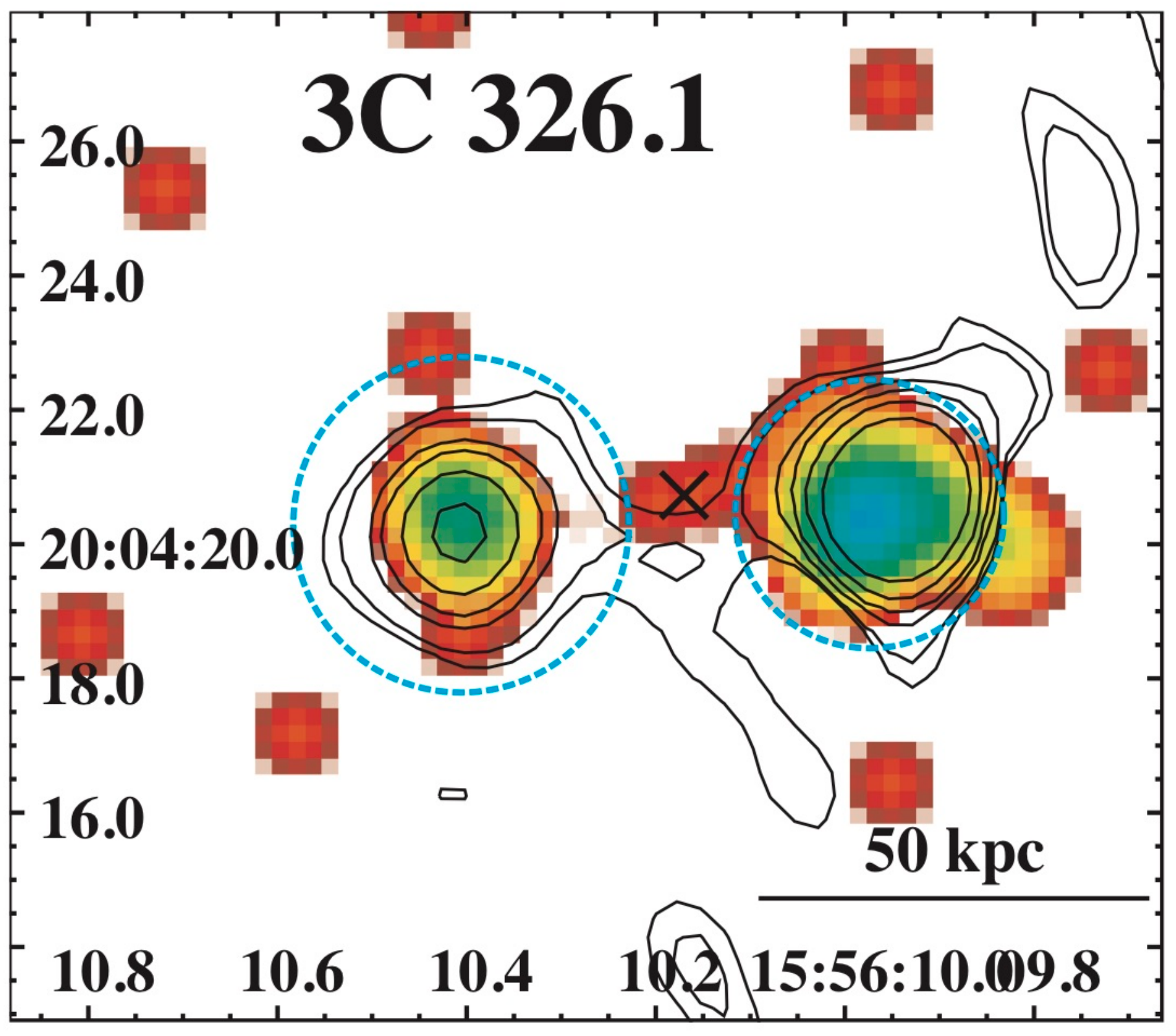}
\caption{0.5 - 7 keV $Chandra$ image with 4.8 GHz VLA contours overlaid. $Chandra$ image was rebinned by a factor of 1/2 and smoothed with a 6 pixel Gaussian kernel. Radio contours were drawn at levels of 5, 10, 30, 50, 100 and 200 times the rms. Radio map has a beam size of 1.5 \arcsec, which is equivalent to 12.9 kpc at the source redshift. The position of the X-ray nucleus is shown with a black X. Blue circles show regions used to identify lobes.}
\label{fig:3c326.1C}
\end{figure}

We defined the position of the X-ray nucleus of this source using the 5 - 7 keV band emission, in which the nucleus was detected with a 2$\sigma$ confidence level. We were not able to detect the X-ray nucleus above a 3$\sigma$ level of confidence, using the full band X-ray emission. We detected both lobes (see Fig. \ref{fig:3c326.1C}) with a confidence level of, at least, 4$\sigma$. As for 3C\,322, we did not detect diffuse emission that could indicate the presence of ICM to corroborate \citet{Kotyla2016} results.

\textit{3C\,418}. 
This is a QSO at $z=1.686$ (\citealt{Veron2006}). According to \citet{Hilbert2016} in the IR this source can be seen as a bright target submerged in a dense field, due to its location at a low galactic latitude; while in the 4.86 GHz VLA map the emission comes from the core and a small jet extending 6 \arcsec\ to the northwest. 

\begin{figure}[h!]
    \centering
\includegraphics[width=8.9cm]{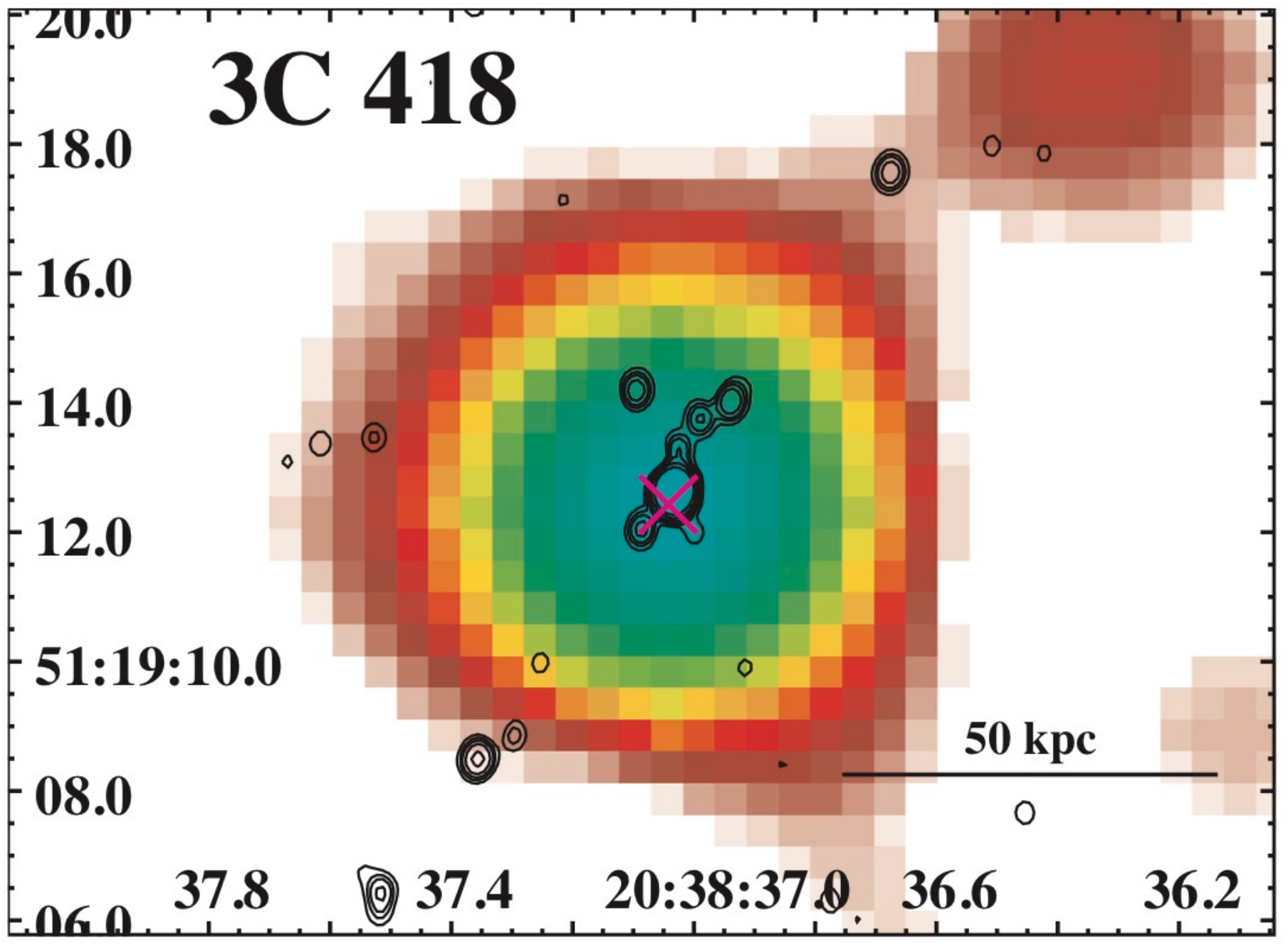}
\caption{0.5 - 7 keV $Chandra$ image with 4.8 GHz VLA contours overlaid. $Chandra$ image was smoothed with a 6 pixel Gaussian kernel. Radio contours were drawn at levels of 10, 20, 30, 50, 100 and 200 times the rms. Radio map has a beam size of 0.4 \arcsec, which is equivalent to 3.4 kpc at the source redshift. The position of the radio core is shown with a magenta X.}
\label{fig:3c418C}
\end{figure}

Using the position of the radio core, we detected the X-ray nucleus in the 5 - 7 keV band above a 5$\sigma$ confidence level. We also detected the nucleus in the full X-ray band and it was the brightest nucleus in the full X-ray band in our sample (see Fig. \ref{fig:3c418C}). Due to the small angular size of the radio emission of 3C\,418 ($\sim3\arcsec$), we did not attempt to detect X-ray counterparts of radio lobes or diffuse X-ray emission in the soft band. The X-ray nucleus of 3C\,418 is affected by 10\% pileup.

\begin{figure}[h!]
    \centering
\includegraphics[width=8.9cm]{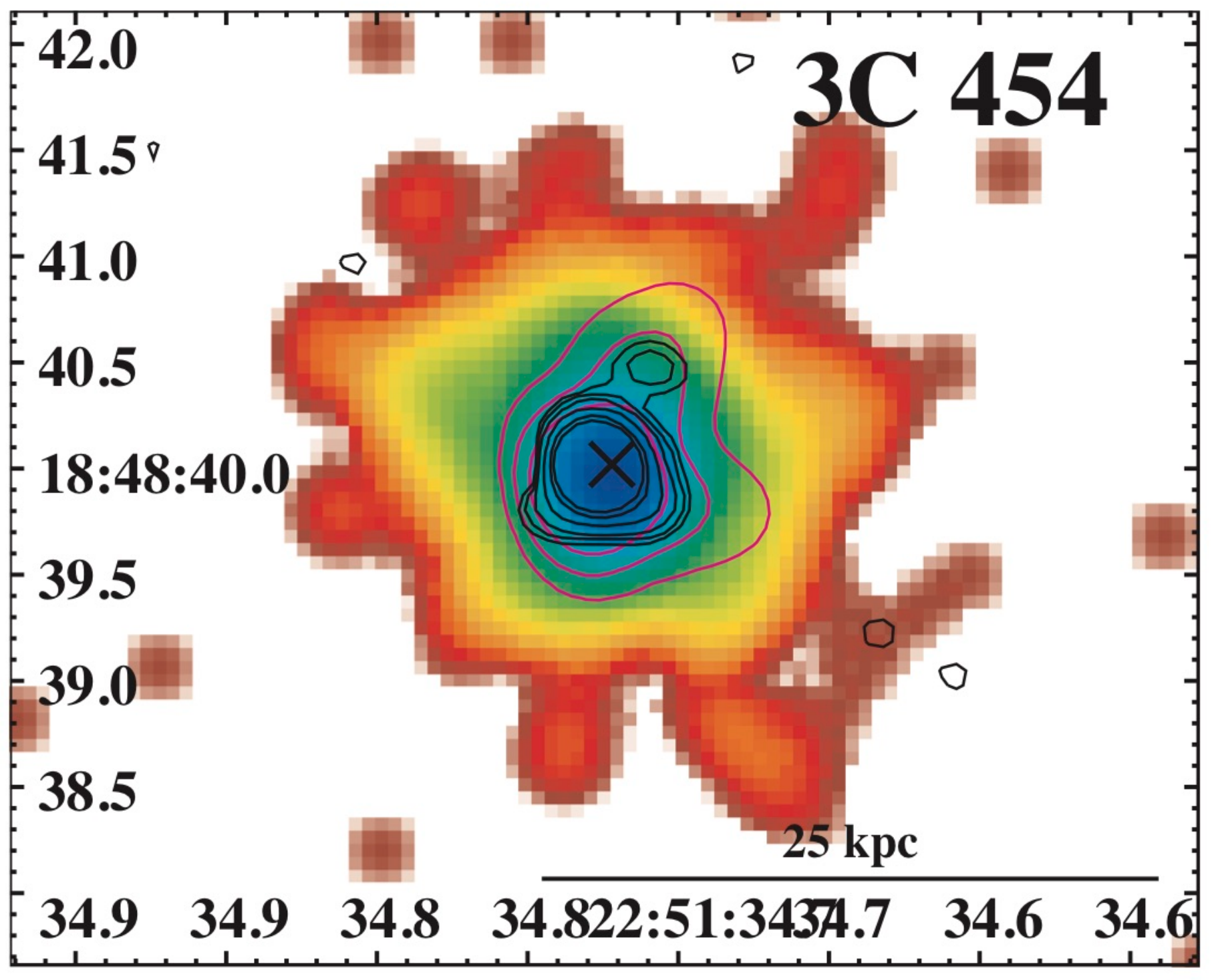}
\caption{0.5 - 7 keV $Chandra$ image with 4.8 GHz (magenta) and 8.4 GHz (black) VLA contours overlaid. $Chandra$ image was rebinned by a factor of 1/8 and smoothed with a 6 pixel Gaussian kernel. 4.8 GHz radio contours were drawn at levels of 20, 100 and 300 times the rms and the 4.8 GHz radio map has a beam size of 0.4 \arcsec, which is equivalent to 3.4 kpc at the source redshift. 8.4 GHz radio contours were drawn at levels of 20, 30, 50, 100, 200 and 300 times the rms and the 8.4 GHz radio map has a beam size of 0.3 \arcsec, which is equivalent to 2.6 kpc at the source redshift. The position of the radio core is shown with a black X.}
\label{fig:3c454X}
\end{figure}

\textit{3C\,454}. 
This source is a QSO at $z=1.757$ (\citealt{Veron2006}) defined as a Compact Steep Spectrum (CSS) source by \citet{ODea1998}. Using the 0.5 - 7 keV $Chandra$ image, we detected the nucleus above a 3$\sigma$ confidence level (see Fig. \ref{fig:3c454X}). The position of this nucleus was given by the position of the radio core, in which the nucleus in the 5 - 7 keV was detected with a confidence level above 5$\sigma$. Radio lobes are not resolved in the radio map and, due to the small angular size of the radio emission of 3C\,454 ($\sim1.5\arcsec$), we did not attempt to detect any extended X-ray counterparts of radio lobes or diffuse X-ray emission in the 0.5 - 3 keV. The X-ray nucleus of 3C\,454 is affected by 3\% pileup.

\textit{3C\,454.1}.
This is a FR II radio galaxy at $z=1.841$ classified as a HERG by \citet{Stickel1996} and as a CSS by \citet{ODea1998}. Using optical observations, \citet{Chiaberge2015} claimed that this source recently underwent merger activity. Lastly, \citet{Kotyla2016} found an overdensity of sources in its field.

\begin{figure}[h!]
    \centering
\includegraphics[width=8.9cm]{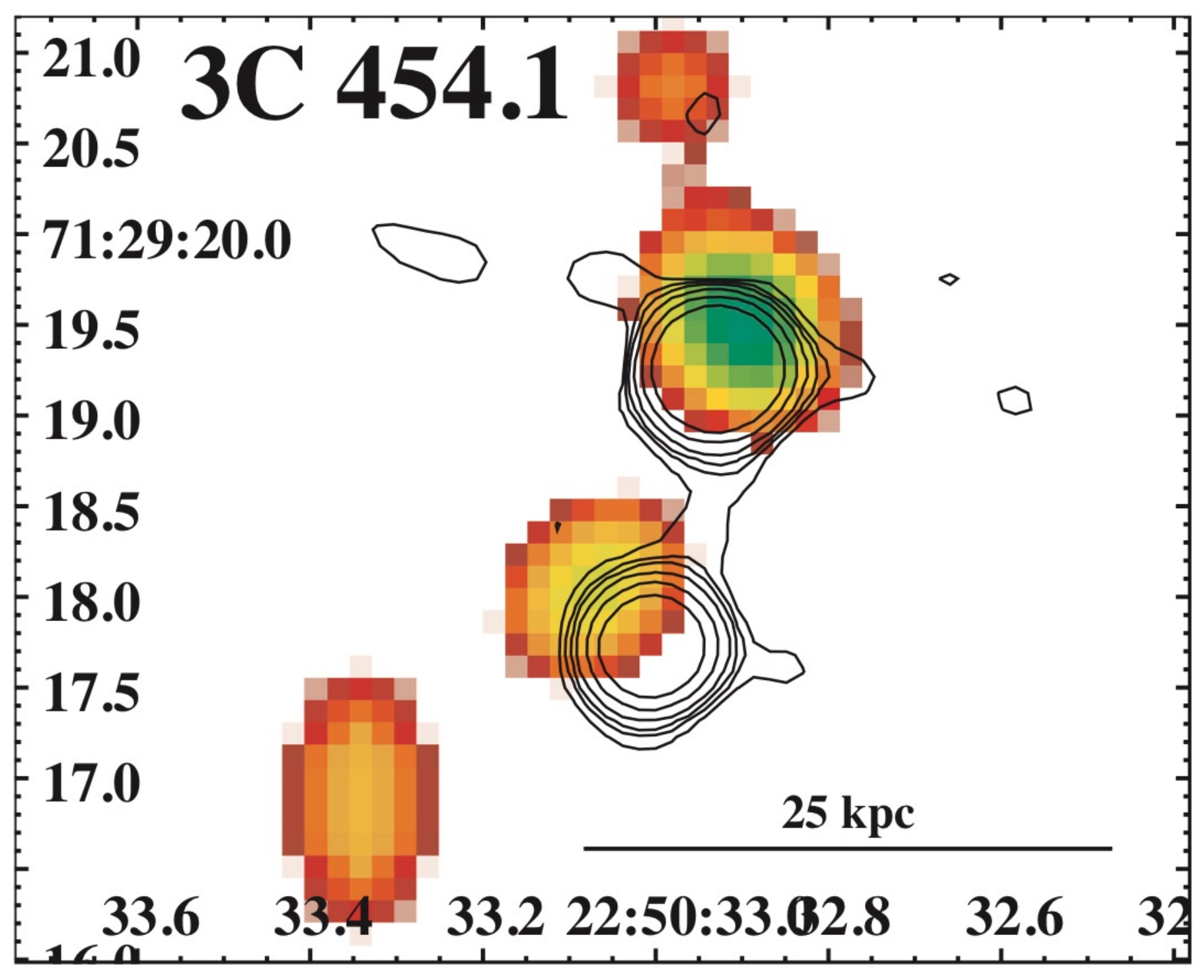}
\caption{0.5 - 7 keV $Chandra$ image with 4.8 GHz VLA contours overlaid. $Chandra$ image was rebinned by a factor of 1/4 and smoothed with a 6 pixel Gaussian kernel. Radio contours were drawn at levels of 10, 20, 30, 50, 100 and 200 times the rms. Radio map has a beam size of 0.4 \arcsec, which is equivalent to 3.4 kpc at the source redshift.}
\label{fig:3c454.1C}
\end{figure}

In the case of 3C\,454.1, we were not able to identify the X-ray nucleus (see Fig. \ref{fig:3c454.1C}). The position reported by \citet{Spinrad1985} matches the northern X-ray peak, however, this peak is most likely the X-ray counterpart of the northern radio lobe. On the other hand, the position of the X-ray peak in the 5 - 7 keV band matches the position of the southern radio lobe and, therefore we concluded that we could not detect the X-ray nucleus of 3C\,454.1. We did not attempt to detect X-ray counterparts of radio lobes or diffuse X-ray emission in the soft band, due to the small angular size of the radio emission of 3C\,454.1 ($\sim3\arcsec$).

\section{Summary and Conclusions}
\label{sec:summary}

We present the X-ray analyses carried out on the 9 remaining identified 3C sources up to $z=2.5$ (namely 3C\,239, 3C\,249, 3C\,257, 3C\,280.1, 3C\,322, 3C\,326.1, 3C\,418, 3C\,454 and 3C\,454.1) observed by $Chandra$ during Cycle 20 as part of the 3CR $Chandra$ Snapshot Survey. This sample is particularly interesting since it covers the highest redshift identified extragalactic sources in the 3CR catalog and, therefore, these are the highest redshift sources observed so far during the 3CR $Chandra$ Snapshot Survey. However, the high redshift prevents us from resolving hotspot emission from lobe emission and, therefore, we label as lobes all extended structures not being nuclei. 
All data presented in this paper, as well as all previous data from the 3CR $Chandra$ Snapshot Survey are publicly available in the $Chandra$ archive.

In this paper, we present the main parameters of these newly observed sources. We followed the same data reduction and analysis procedures as in previous papers of the 3CR $Chandra$ Snapshot Survey. We created flux maps in the soft (0.5 - 1 keV), medium (1 - 2 keV) and hard (2 - 7 keV) bands for detected nuclei and extended X-ray features (i.e., lobes) and reported their background-subtracted number of photons, detection significance, fluxes in each band and X-ray luminosities.

We detected the X-ray nuclei using the 0.5 - 7 keV emission in all sources in our sample except for 3C\,326.1 and 3C\,454.1. Additionally, we detected X-ray counterparts of radio lobes in the same band in 6 of the 9 sources in our sample, namely 3C\,239, 3C\,249, 3C\,257, 3C\,280.1, 3C\,322 and 3C\,326.1. In particular, we detected both lobes in 3C\,249, 3C\,280.1 and 3C\,326.1. In the three sources left, the X-ray detected lobe was also the brightest in radio. Lastly, we found diffuse X-ray emission along the radio axis of 3C\,249 in the 0.5 - 3 keV band.

This paper marks the completion of the $Chandra$ Snapshot Survey of identified 3CR sources, carried out to guarantee the X-ray coverage of the entire 3CR extragalactic catalog with $Chandra$. Thanks to these last 9 observations presented here we completed the analysis of all identified sources listed in the 3CR. 

From providing a first glimpse into interesting sources that warranted deeper observations, to providing data for more detailed studies, this survey has proven to be an extremely helpful tool in the study of radio sources (see e.g., \citealt{Hardcastle2010,Hardcastle2012}, \citealt{Balmaverde2012}, \citealt{Orienti2012}, \citealt{Ineson2013}, \citealt{Dasadia2016} and \citealt{Madrid2018}). So far, during the 3CR $Chandra$ Snapshot Survey, we have observed a total of 122 sources that had no previous observations in the $Chandra$ archive. Using these observations, we have been able to detect X-ray counterparts of 119 radio cores.

According to the XJET database\footnote{http://hea-www.harvard.edu/XJET/}, thanks to the 3CR $Chandra$ Snapshot Survey, we discovered 8 more radio jet knots with an X-ray counterpart in 7 different sources thus increasing their number by $\sim$10\% (see \citealt{Massaro2011} for latest results on the XJET project). In addition, we also doubled the number of sources having at least one X-ray detected hotspot. Then, in all previous data papers of the 3CR $Chandra$ Snapshot Survey, we reported the discovery of the X-ray counterpart of radio lobes for 11 3CR sources in addition to those listed in the literature. Additionally, X-ray emission arising from ICM in galaxy clusters was detected in 19 sources including those reported in the archival search (\citealt{Massaro2015}). Having such a large sample of X-ray counterparts of radio galaxies' components will allow us to study the origin of the X-ray emission from jets and hotspots which is currently still uncertain, but believed to be non-thermal (see e.g., \citealt{Harris2002,Harris2006} and \citealt{Worrall2009}), the most likely scenarios being synchrotron emission or inverse Compton (IC) scattering.

We remark that the current summary about the X-ray detection of extended radio structures in 3CR sources does not include the following cases for which extensive analyses have been carried out in the literature, namely: 3CR\,66A, 3CR\,71 (a.k.a. NGC\,1068), 3CR\,84 (a.k.a Perseus A), 3CR\,186, 3CR\,231 (a.k.a. M82), 3CR\,317 (a.k.a. Abell 2052) and 3CR\,348 (a.k.a. Hercules A) as well as 3CR\,236, 3CR\,263 and 3CR\,386 for which the multifrequency analysis is still ongoing (Birkinshaw priv. comm.).

\begin{figure}
    \centering
\includegraphics[width=8.9cm]{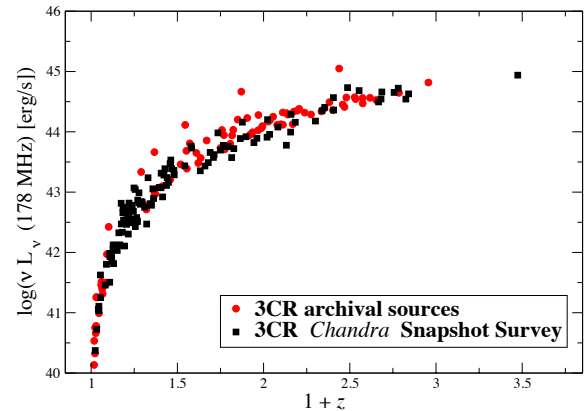}
\caption{Radio luminosity at 178 MHz vs. redshift plot for the 3CR sources with observations in the $Chandra$ archive (red circles) and those observed during the 3CR $Chandra$ Snapshot Survey (black squares).}
\label{fig:lz}
\end{figure}

\begin{figure*}
    \centering
\includegraphics[width=8.9cm]{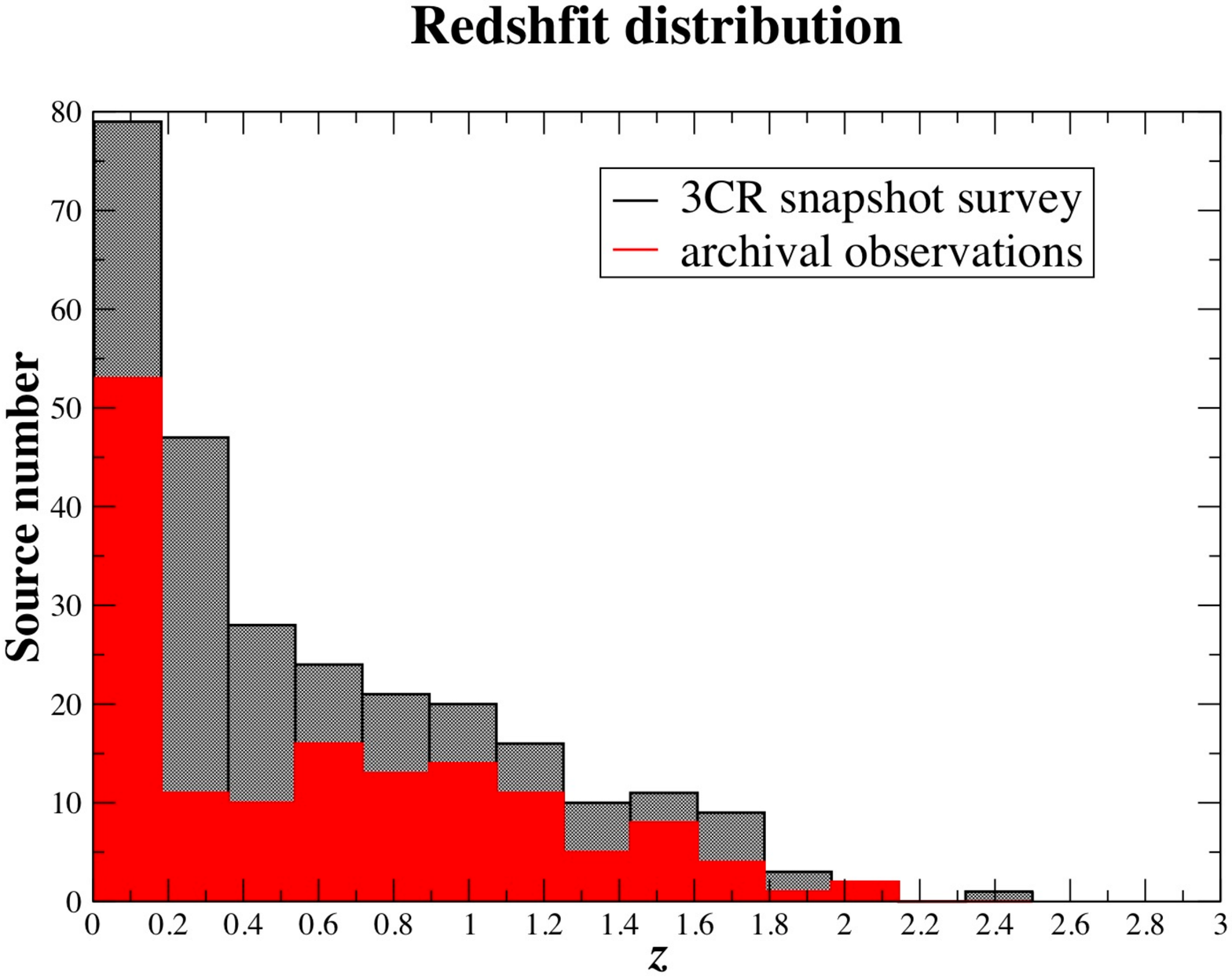}
\includegraphics[width=8.9cm]{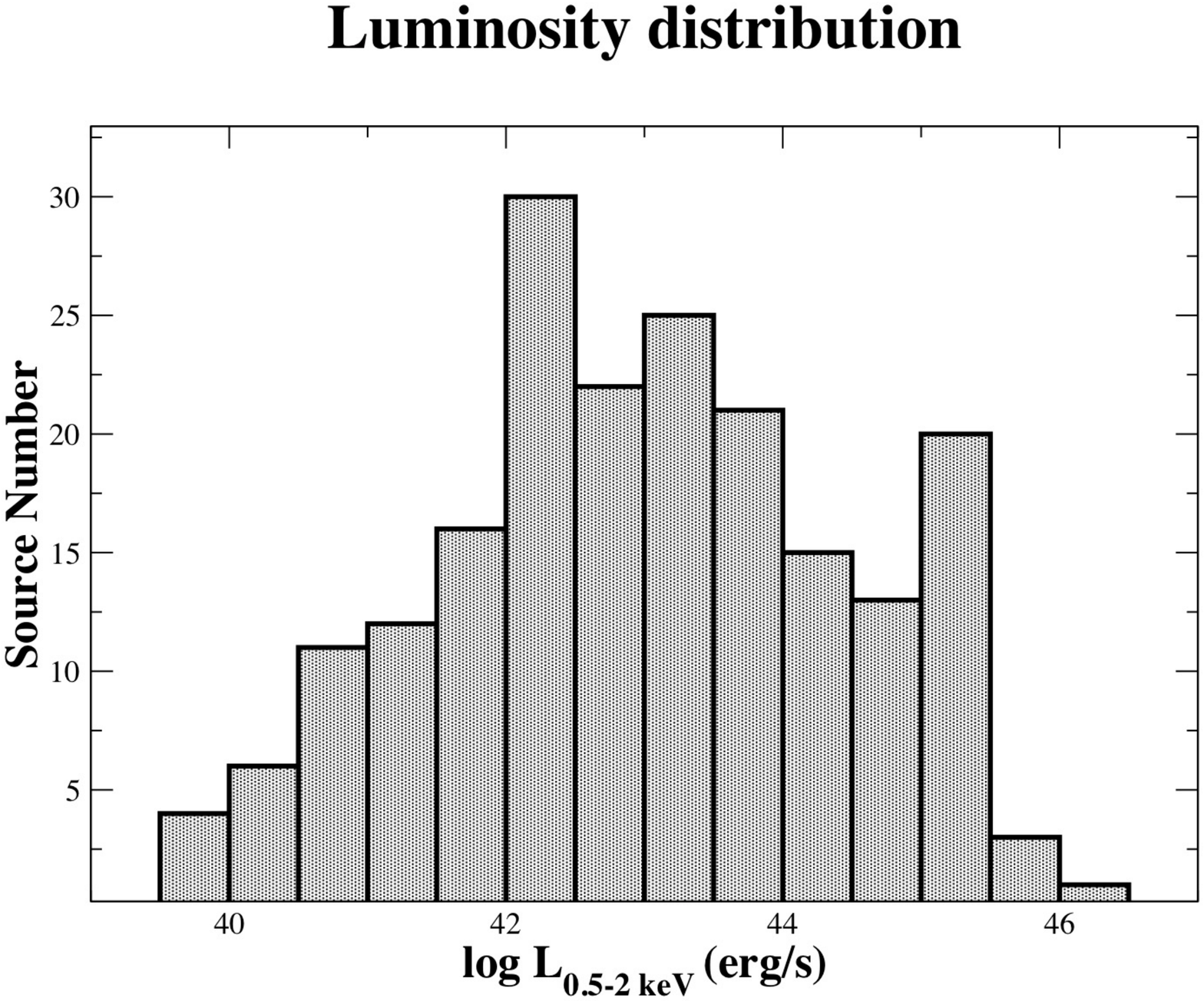}
\caption{Left: Redshift distribution of sources in the 3CR $Chandra$ Snapshot Survey (in grey) and in the $Chandra$ archive before the Snapshot Survey (red). Right: Soft X-ray luminosity distribution of 3CR nuclei in the $Chandra$ archive.}
\label{fig:distributions}
\end{figure*}

In Fig. \ref{fig:lz}, we show the radio luminosity at 178 MHz obtained form \citet{Spinrad1985} versus redshift for 3CR sources with $Chandra$ observations, comparing those we observed during the 3CR $Chandra$ Snapshot Survey and those with previous archival observations. This comparison shows that at redshift $z>0.5$ our snapshot observations pointed the faintest radio sources. In the left panel of Fig. \ref{fig:distributions}, we show the redshift distribution of the 3CR sources before and after our survey was completed, including all sources analyzed in the current manuscript, and not considering those still unidentified (i.e., lacking an optical counterpart and thus a $z$ estimate). The comparison between these two $z$ distributions highlights the impact of our 3CR $Chandra$ Snapshot Survey on the number of sources at moderate redshift (i.e., above 1.2) as well as in the low redshift range (i.e., between 0.1 and 0.5) which were significantly increased with respect to those listed in the $Chandra$ archive. Finally, the right panel of Fig. \ref{fig:distributions} shows the distribution of the X-ray nucleus luminosities as measured in all data papers published to date (\citealt{Massaro2010,Massaro2012,Massaro2013,Massaro2015,Massaro2018} and \citealt{Stuardi2018}) and in this paper.

To complete the 3CR $Chandra$ Snapshot Survey, the next step will be to obtain observations of the 25 3C sources that remain unidentified. A $Chandra$ campaign is currently ongoing to observe the first 9 unidentified sources and results will be presented in a forthcoming paper. Additionally, we are planning another forthcoming paper presenting a statistical study on the whole 3CR $Chandra$ Snapshot Survey in which we plan to include the SED modelling of these sources.

\acknowledgments 
We thank the anonymous referee for useful comments that led to improvements in the paper.
We are grateful to M. Hardcastle and C. C. Cheung for providing
several radio images of the 3CR sources while the remaining ones were
downloaded from the NVAS\footnote{http://archive.nrao.edu/nvas/} (NRAO
VLA Archive Survey), NED\footnote{http://ned.ipac.caltech.edu/} (Nasa
Extragalactic Database) and from the DRAGN
webpage\footnote{http://www.jb.man.ac.uk/atlas/}.
This investigation is supported by the National Aeronautics and Space Administration (NASA) grants GO6-17081X and GO9-20083X. The Chandra X-ray Observatory Center is operated by the Smithsonian Astrophysical Observatory for and on behalf of NASA under contract NAS8-03060. This work is also supported by the ``Departments of Excellence 2018 - 2022'' Grant awarded by the Italian Ministry of Education, University and Research (MIUR) (L. 232/2016). This research has made use of resources provided by the Compagnia di San Paolo for the grant awarded on the BLENV project (S1618\_L1\_MASF\_01) and by the Ministry of Education, Universities and Research for the grant MASF\_FFABR\_17\_01. 
A.J. acknowledges the financial support (MASF\_CONTR\_FIN\_18\_01) from the Italian National Institute of Astrophysics under the agreement with the Instituto de Astrofisica de Canarias for the ``Becas Internacionales para Licenciados y/o Graduados Convocatoria de 2017''.
F.M. acknowledges financial contribution from the agreement ASI-INAF n.2017-14-H.0.
A.P. acknowledges financial support from the Consorzio Interuniversitario per la Fisica Spaziale (CIFS) under the agreement related to the grant MASF\_CONTR\_FIN\_18\_02.
C.S. acknowledges support from the ERC-StG DRANOEL, n. 714245.
F.R. acknowledges support from FONDECYT Postdoctorado 3180506 and CONICYT project Basal AFB-170002.
C.O. and S.B. acknowledge support from the Natural Sciences and Engineering Research Council (NSERC) of Canada.

The National Radio Astronomy Observatory is operated by Associated Universities, Inc.,
under contract with the National Science Foundation.
This research has made use of data obtained from the High-Energy Astrophysics Science Archive
Research Center (HEASARC) provided by NASA's Goddard Space Flight Center; 
the SIMBAD database operated at CDS,
Strasbourg, France; the NASA/IPAC Extragalactic Database
(NED) operated by the Jet Propulsion Laboratory, California
Institute of Technology, under contract with the National Aeronautics and Space Administration.
TOPCAT\footnote{\underline{http://www.star.bris.ac.uk/$\sim$mbt/topcat/}} 
\citep{Taylor2005} for the preparation and manipulation of the tabular data and the images.
SAOImage DS9 were used extensively in this work
for the preparation and manipulation of the
images.  SAOImage DS9 was developed by the Smithsonian Astrophysical
Observatory. 

{Facilities:} \facility{VLA}, \facility{MERLIN}, \facility{CXO (ACIS)}, \facility{GMRT}

\clearpage
\begin{table*} 
\caption{Source List of the \chn\ AO20 Snapshot survey}
\label{tab:log}
\begin{center}
\footnotesize
\scriptsize
\begin{tabular}{|llllllllllll|}
\hline
3CR  & Class & R.A. (J2000) & Dec. (J2000) & z & kpc scale    & D$_L$ & log(N$_{H,Gal}$) & m$_v$ & S$_{178}$ &\chn\   & Obs. Date  \\
name &          & (hh mm ss)   & (dd mm ss)    &   & (kpc/arcsec) & (Mpc)  & (cm$^{-2}$)   &            & (Jy)            & Obs. ID & yyyy-mm-dd \\ 
\hline 
\noalign{\smallskip}
  239 & RG - FR\,II - HERG & 10 11 45.284 & +46 28 18.79 & 1.781 & 8.598 & 13715.2 & 19.84 & 22.5 & 13.2 & 21395 & 2019-02-20 \\
  249 & QSO - LDQ & 11 02 03.774 & -01 16 16.67 & 1.554 & 8.612 & 11587.3 & 20.60 & - & 16.9 & 21396 & 2019-02-16\\
  257 & RG & 11 23 09.391 & +05 30 18.50 & 2.474 &  8.245 & 20524.7 & 20.47 & - & 9.7 & 21397 & 2019-02-13\\
  280.1 & QSO - LDQ & 13 00 33.364 & +40 09 07.28 & 1.667 & 8.615 & 12639.0 & 20.21 & 19.4 & 9.2 & 21398 & 2019-07-29\\
  322 & RG - FR\,II & 15 35 01.269 & +55 36 52.33 & 1.681 & 8.614 & 12770.4 & 20.11 & 23.0 & 10.1 & 21399 & 2019-02-25\\
  326.1 & RG & 15 56 10.170 & +20 04 20.73 & 1.825 & 8.586 & 14134.4 & 20.51 & R & 8.2 & 21400 & 2019-05-20\\
  418 & QSO & 20 38 37.042 & +51 19 12.43 & 1.686 & 8.613 & 12817.4 & 21.72 & 20.0 & 13.1 & 21401 & 2018-09-03\\
  454 & QSO - CSS & 22 51 34.722 & +18 48 40.02 & 1.757 & 8.603 & 13487.5 & 20.75 & 18.5 & 11.6 & 21403 & 2019-08-19\\
  454.1 & RG - FR\,II - CSS & 22 50 33.067 & +71 29 18.02 & 1.841 & 8.582 & 14287.3 & 21.45 & - & 9.8 & 21402 & 2019-03-16\\
\noalign{\smallskip}
\hline
\end{tabular}\\
\end{center}
Col. (1): The 3CR name.
Col. (2): The `Class' column specifies whether the source is a radio galaxy (RG) a quasar (QSO) and the source classification according to the radio morphology (only Fanaroff-Riley class II, FR II, in our case for radio galaxies; \citealt{Fanaroff1974}; or lobe-dominated quasar, LDQ), the optical spectroscopic designation (HERG, ``High Excitation Radio Galaxy'') and whether the source is classified as Compact Steep Spectrum (CSS) (see also \citealt{perryman84,hes96,grimes04} for more details). These classifications were listed when available in the literature.
Col. (3-4): The celestial positions listed are those of the nuclei: Right ascension and Declination (equinox J2000, see \S~\ref{sec:obs} for details). We reported here the X-ray peak in the 5 - 7 keV band, which corresponds to the X-ray nucleus position, for sources for which the radio core was not clearly detected, which were all except 3C\,280.1, 3C\,418 and 3C\,454.
Col. (5): Redshift $z$. We verified in the literature (e.g., NED and/or SIMBAD databases) if new $z$ values were reported after the release of the 3CR catalog. The only sources with the original \citet{Spinrad1985} are 3C\,239, 3C\,418 and 3C\,454; while redshifts of 3C\,257, 3C\,322, 3C\,326.1 and 3C\,454.1 were obtained from \citet{Hewitt1991}. Lastly, redshifts of 3C\,249 and 3C\,280.1 were obtained from \citet{Best2003} and \citet{Hewett2010}, respectively.
Col. (6): The angular to linear scale factor in arcseconds. Cosmological parameters used to compute it are reported in \S~\ref{sec:intro}.
Col. (7): Luminosity Distance in Mpc. Cosmological parameters used to compute it are reported in \S~\ref{sec:intro}. 
Col. (8): Galactic Neutral hydrogen column densities N$_{H,Gal}$ along the line of sight (\citealt{kalberla05}).
Col. (8): The optical magnitude in the V band listed in the 3CR catalog (\citealt{Spinrad1985}).
Col. (9): S$_{178}$ is the flux density at 178 MHz (\citealt{Spinrad1985}).
Col. (10): The \chn\ observation ID.
Col. (11): The date when the \chn\ observation was performed.\\
\end{table*}


\begin{table*} 
\caption{X-ray emission from nuclei.}
\label{tab:cores}
\begin{center}
\scriptsize
\begin{tabular}{|cccccccccc|}
\hline
3CR  & Net & Detection & F$_{0.5-1~keV}^*$ & F$_{1-2~keV}^*$ & F$_{2-7~keV}^*$ & F$_{0.5-7~keV}^*$ & log(N$_{H}(z)$)  & L$_X$ & L$_X,obs$ \\
name & counts & Significance & (cgs)                 & (cgs)           & (cgs)           & (cgs)  &  (cm$^{-2}$)  & (10$^{44}$erg~s$^{-1}$) & (10$^{44}$erg~s$^{-1}$) \\
\hline 
\noalign{\smallskip}
239 & 7.5 (2.7) & $>$5 & -- & 0.6 (0.5) & 4.4 (1.9) & 5.0 (2.0) & 22.863 - 23.845 & 0.283 - 0.940 & 0.189 - 0.300 \\
249 & 11.5 (3.4) & $>$5 & 1.01 (1.08) & 0.6 (0.4) & 4.8 (1.8) & 6.4 (2.1) & 19.89 - 22.38 & 0.234 - 0.304 & 0.222 - 0.233\\
257 & 29.5 (5.4) & $>$5 & -- & 3.18 (1.02) & 15.5 (3.6) & 18.7 (3.8) & 23.412 - 23.813 & 3.192 - 6.087 & 1.791 -  2.528 \\
280.1$^{**}$ & 719.6 (26.8) & $>$5 & 91.4 (11.8) & 123.0 (6.5) & 251.7 (14.5) & 466.2 (19.8) & 20.01 - 21.507 & 17.675 - 17.914 & 16.607 - 17.482\\
322 & 3.6 (1.9) & 4 & -- & -- & 3.6 (1.9) & 3.6 (1.9) & 19.991 - 24.001 & 0.090 - 0.839 & 0.088 - 0.207\\
326.1$^{+}$ & 1.6 (1.3) & -- & -- & -- & 1.8 (1.4) & 1.8 (1.4) & 24.4 - 24.404 & 0.800 - 1.040 & 0.110 - 0.144 \\
418$^{***}$ & 1513.6 (38.9) & $>$5 & 22.4 (4.7) & 166.0 (7.2) & 784.8 (25.4) & 973.3 (26.8) & 22.871 - 23.023 & 69.445 - 73.719 & 39.229 - 39.418\\
454$^{**}$ & 438.7 (20.9) & $>$5 & 50.7 (8.8) & 70.8 (4.9) & 155.2 (11.1) & 276.7 (15.0) & 19.902 - 22.338 & 12.269 - 13.608 & 10.728 - 11.388\\
454.1 & -- & -- & -- & -- & -- & -- & --  & -- & --  \\

\noalign{\smallskip}
\hline
\end{tabular}\\
\end{center}
Col. (1): 3CR name.
Col. (2): Background-subtracted number of photons within a circle of radius $r=2$\arcsec\ in the 0.5 - 7 keV band. Number between parenthesis corresponds to the uncertainty in the number of photons computed assuming Poisson statistics.
Col. (3): Detection significance of the nucleus in the 0.5 - 7 keV band, assuming the background has a Poisson distribution with a mean scaled to a 2\arcsec\ radius circle.
Col. (4): Measured X-ray flux between 0.5 and 1 keV.
Col. (5): Measured X-ray flux between 1 and 2 keV.
Col. (6): Measured X-ray flux between 2 and 7 keV.
Col. (7): Measured X-ray flux between 0.5 and 7 keV.
Col. (8):Range of intrinsic column densities needed to obtain the X-ray fluxes reported assuming a model comprising Galactic absorption (fixed to the values reported in \ref{tab:log}), a power-law with slope fixed to 1.8, and intrinsic absorption with column density N$_{H,int}$ at the source redshift $z$.
Col. (9): Range of X-ray luminosities in the 0.5 - 7 keV band without taking into account intrinsic or Galactic absorption.
Col. (10):Range of observed X-ray luminosities in the 0.5 - 7 keV band evaluated from the model with Galactic and intrinsic absorption.\\
Note:\\
($^*$) Fluxes are given in units of 10$^{-15}$erg~cm$^{-2}$s$^{-1}$ and 1$\sigma$ uncertainties are given in parenthesis. 
The uncertainties on the flux measurements are computed as described in \S~\ref{sec:obs}. Fluxes were not corrected for Galactic absorption and were obtained by assuming a flat energy response in each band.\\

($^{**}$) For 3CR\,280.1 and 3CR\,454 the number of photons allowed us to extract their spectra and we obtained acceptable fits (i.e., reduced $\chi^2$ $\simeq$ 0.9) adopting an absorbed power-law model, with only Galactic column density. We found the best fit value of the photon index being 1.54 (0.09) and 1.59 (0.07) with an integrated flux equal to 3.19$\cdot10^{-13}$  erg/cm$^2$/s and 5.51$\cdot10^{-13}$ erg/cm$^2$/s for 3CR\,280.1 and 3CR\,454, respectively. Then, assuming a fixed value of the photon index of 1.8 we also estimated the value of the intrinsic absorption (i.e., $N_H,int$) consistent with those obtained with the photometric analysis (i.e., hardness ratios) within 1$\sigma$ being 1.84(0.07)$\cdot 10^{-22}$ and 0.97(0.04)$\cdot 10^{-22}$ for 3CR\,280.1 and 3CR\,454, respectively.\\

($^{***}$) For 3CR\,418 we also carried out a spectral analysis and fitting with an absorbed power-law model, both including the Galactic and the instrinsic absorption, with fixed photon index of 1.8, we obtained an estimate of the pile up fraction of 0.13 (0.03) and a value of $N_{H,int}$ equal to 4.52$\cdot10^{-22}$ cm$^{-2}$ consistent with what expected from the source count rate using PIMMS v4.10\footnote{https://cxc.harvard.edu/toolkit/pimms.jsp} and that of the $pileup\_map$ task. All consistent with the photometric analysis.\\

($^{+}$) For 3CR\,326.1 we report the upper limit of the intrinsic column density since the nucleus was not detected.
\end{table*}

\begin{table*} 
\caption{X-ray emission from radio extended structures (i.e., lobes).}
\label{tab:features}
\begin{center}
\begin{tabular}{|cccccccccc|}
\hline
3CR  & Component & Size & Net & Detection & F$_{0.5-1~keV}^*$ & F$_{1-2~keV}^*$ & F$_{2-7~keV}^*$ & F$_{0.5-7~keV}^*$ & L$_X$ \\
name &        & (arcsec) & counts & Significance & (cgs)                 & (cgs)           & (cgs)           & (cgs)             & (10$^{44}$erg~s$^{-1}$) \\
\hline 
\noalign{\smallskip}
239  & e4  & 3 & 7.9 (2.8) & 5    & 2.1 (1.7) & 1.1 (0.6) & 1.5 (1.1) & 4.7 (2.1) & 1.0 (0.5)\\
249  & n11 & 4 & 4.1 (2.0) & 3    & 0.7 (1.0) & 0.5 (0.5) & 1.8 (1.8) & 3.0 (2.1) & 0.5 (0.3)\\
249  & s12 & 2.5 & 9.3 (3.0) & $>$5 & 2.0 (1.5) & 1.9 (0.7) & 0.4 (0.8) & 4.2 (1.8) & 0.7 (0.3)\\
257  & w6  & 2 & 8.5 (2.9) & $>$5 & --        & 0.9 (0.6) & 3.9 (1.7) & 4.9 (1.8) & 2.5 (0.9)\\
280.1& w12 & 4.5 & 7.8 (2.8) & 4    & --        & 1.4 (0.8) & 4.6 (2.5) & 6.1 (2.6) & 1.2 (0.5)\\
280.1& e7  & 5 & 26.0 (5.1)& $>$5 & 3.6 (2.3) & 2.3 (0.9) & 15.5 (4.1)& 21.4 (4.8)& 4.1 (0.9)\\
322  & n14 & 3.5 & 7.7 (2.8) & 5    & 0.9 (1.1) & 1.4 (0.7) & 1.9 (1.5) & 4.2 (2.0) & 0.8 (0.4)\\
326.1& e3  & 2.5 & 4.3 (2.1) & 4    & 2.3 (1.7) & 0.2 (0.2) & 1.3 (1.2) & 3.8 (2.1) & 0.9 (0.5)\\
326.1& w3  & 2 & 7.6 (2.8) & $>$5 & 2.8 (2.0) & 1.4 (0.7) & 0.7 (1.0) & 4.9 (2.3) & 1.2 (0.5)\\

\noalign{\smallskip}
\hline
\end{tabular}\\
\end{center}
Col. (1): 3CR name.
Col. (2): Component name is a combination of one letter indicating the orientation of the radio structure and one number indicating distance from the nucleus in arcseconds.
Col. (3): Radius of the circle containing the 3 rms noise contours of the radio emission.
Col. (4): Background-subtracted number of photons within the photometric circle in the 0.5 - 7 keV band. Number between parenthesis corresponds to the uncertainty in the number of photons computed assuming Poisson statistics.
Col. (5): Detection significance of the lobes in the 0.5 - 7 keV band, assuming the background has a Poisson distribution with a mean scaled to a 2\arcsec\ radius circle.
Col. (6): Measured X-ray flux between 0.5 and 1 keV.
Col. (7): Measured X-ray flux between 1 and 2 keV.
Col. (8): Measured X-ray flux between 2 and 7 keV.
Col. (9): Measured X-ray flux between 0.5 and 7 keV.
Col. (10): X-ray luminosity in the range 0.5 to 7 keV with the 1$\sigma$ uncertainties given in parenthesis.\\
Note:\\
($^*$) Fluxes are given in units of 10$^{-15}$erg~cm$^{-2}$s$^{-1}$ and 1$\sigma$ uncertainties are given in parenthesis.
The uncertainties on the flux measurements were computed as described in \S~\ref{sec:obs}. Fluxes were not corrected for Galactic absorption and were obtained by assuming a flat energy response in each band.\\
\end{table*}

\end{document}